\documentclass{aa}  
\usepackage{graphicx}
\usepackage{natbib}
\usepackage{color}
\usepackage{txfonts}
\def\gr{$\gamma$-ray}
\newcommand{\axp}{\mbox{1E 1547.0-5408}}
\newcommand{\axpp}{\mbox{1E 2259+586}}
\newcommand{\intgr}{{\it INTEGRAL}}

\begin{document}
\authorrunning{Savchenko et al.}
\titlerunning{Exceptional flaring activity of the anomalous X-ray pulsar \axp}

   \title{SGR-like flaring activity of the anomalous X-ray pulsar \axp}

   \author{V.Savchenko\inst{1,2}
          \and
          A.Neronov\inst{1,2}
        \and V.Beckmann\inst{1,2}
          \and N.Produit\inst{1,2}
          \and R.Walter\inst{1,2}
          }

   \institute{     ISDC Data Centre for Astrophysics, Ch. d'Ecogia 16, 1290 Versoix, Switzerland
  \and Geneva Observatory, Ch. des Maillettes 51, 1290 Sauverny, Switzerland       }
   \date{Received ; accepted }

 
  \abstract
{} {We studied an exceptional period of activity of the anomalous X-ray pulsar \axp\ in  January 2009, during which about $200$ hard X-ray / soft \gr\ bursts were detected by different  instruments on board of ESA's \gr\ observatory {\it INTEGRAL}.}{The major activity episode (22 January 2009), which was detected by NASA's \gr\ telescope {\it Swift}, happened when the source was outside the field of view of all the \intgr\ instruments. But we were still able to study the statistical properties  as well as spectral and timing characteristics of 84 short ($100$~ms$-10$~s) bursts from the source during this activity period detected simultaneously by the anti-coincidence shield of the spectrometer SPI on board of \intgr\ and by the detector of the imager IBIS/ISGRI.    }{We find that the luminosity of the 22 January 2009 bursts of \axp\  was $\ge 10^{42}$~erg~s$^{-1}$ above $\sim 80$~keV. This luminosity is comparable 
to that of the bursts of soft gamma repeaters (SGR) and is at least two orders of magnitude larger than the luminosity of the previously reported bursts from AXPs. Similarly to the SGR bursts,  the brightest bursts of \axp\ consist of a short spike of  $\sim 100$~ms duration with a hard spectrum, followed by a softer extended tail of 1-10~s duration, which occasionally exhibits pulsations with the source spin period of $\sim 2$~s. The "short spike + tail" template is not valid for all the observed bursts. We also observe a certain amount of $\sim 1$~s duration bursts either without a spike or with a spike delayed with respect to the onset of the burst. A third type of bursts is the "flat-top" burst of sub-second duration, which is similar to the "precursors" observed in the SGR flares. We find that the bursts of \axp\ harden with increasing luminosity. Such a behavior is opposite of those observed in SGR bursts, but is similar to the hardness-luminosity relation observed in AXP 1E 2259+586. We confirm  the existence of a correlation between the burst fluences and durations and the absence of a correlation between the burst fluence and the waiting time between subsequent bursts, previously established for the AXP \axpp.  }
{The observation of AXP bursts with luminosities comparable to the one of SGR bursts strengthens the conjecture that AXPs and SGRs are different representatives of one and the same source type, and that the AXP and SGR bursts are triggered by a similar type of instabilities.  }


   \maketitle
%

\section{Introduction}

Anomalous X-ray pulsars (AXP) and soft gamma repeaters (SGR) are believed to be young neutron stars  with ultra-strong magnetic fields exceeding the Schwinger magnetic field $B_{\rm Schw}=m_e^2c^3/e\hbar\simeq 4.4\times 10^{13}$~G \citep{duncan92,paczynski92,thompson95,mereghetti08}.

SGRs are known to exhibit periods of activity, during which they emit a large amount of short bursts with typical durations of $\sim 100$~ms and a luminosity reaching $\sim 10^{42}$~erg~s$^{-1}$ (see \citet{mereghetti08} for a recent review). The bursting activity appears to be a random process with a waiting time between the bursts distributed lognormally  and with no correlation between the waiting time and the burst intensity \citep{hurley94}. The burst intensities follow a 
power law distribution \citep{gogus99}.  Similar bursting activity is observed also in AXPs \citep{gavriil02,kaspi03,gavriil04,gavriil07,woods05,israel07}. However, the AXP bursts detected up to now have much lower (by several orders of magnitude) luminosities. Besides, contrary to the SGR bursts, which exhibit spectral softening with increasing luminosity \citep{gogus00}, the brighter AXP bursts appear to have harder spectra \citep{gavriil04}.

The lightcurves of the AXP bursts have a rich morphology. \citet{woods05} introduced a tentative classification for the AXP bursts, dividing them into short bursts with symmetric time profiles (type A) and longer fast-rise / slow-decay bursts with the decaying tails lasting tens to hundred seconds (type B). The "type A" bursts resemble the short bursts of SGRs (although with much lower luminosity), while the "type B" bursts are weak analogs of the so-called "giant flares" of SGRs which are usually characterized by a short spike followed by a longer pulsating tail lasting up to thousands of seconds. It is not clear if such a classification corresponds to physically different types of bursts (e.g. the bursts triggered by the fractures of the neutron star crusts \citep{thompson95} and the bursts triggered by magnetic reconnection events \citep{lyutikov02} and whether it is exhaustive (i.e. all the bursts are either type A or type B). It is also not clear if the dramatically different luminosity of the SGR and AXP bursts reflects the difference of  physical conditions in the two source classes or the difference in the mechanisms leading to the production of the bursts. 

\begin{figure*}
\includegraphics[width=\linewidth]{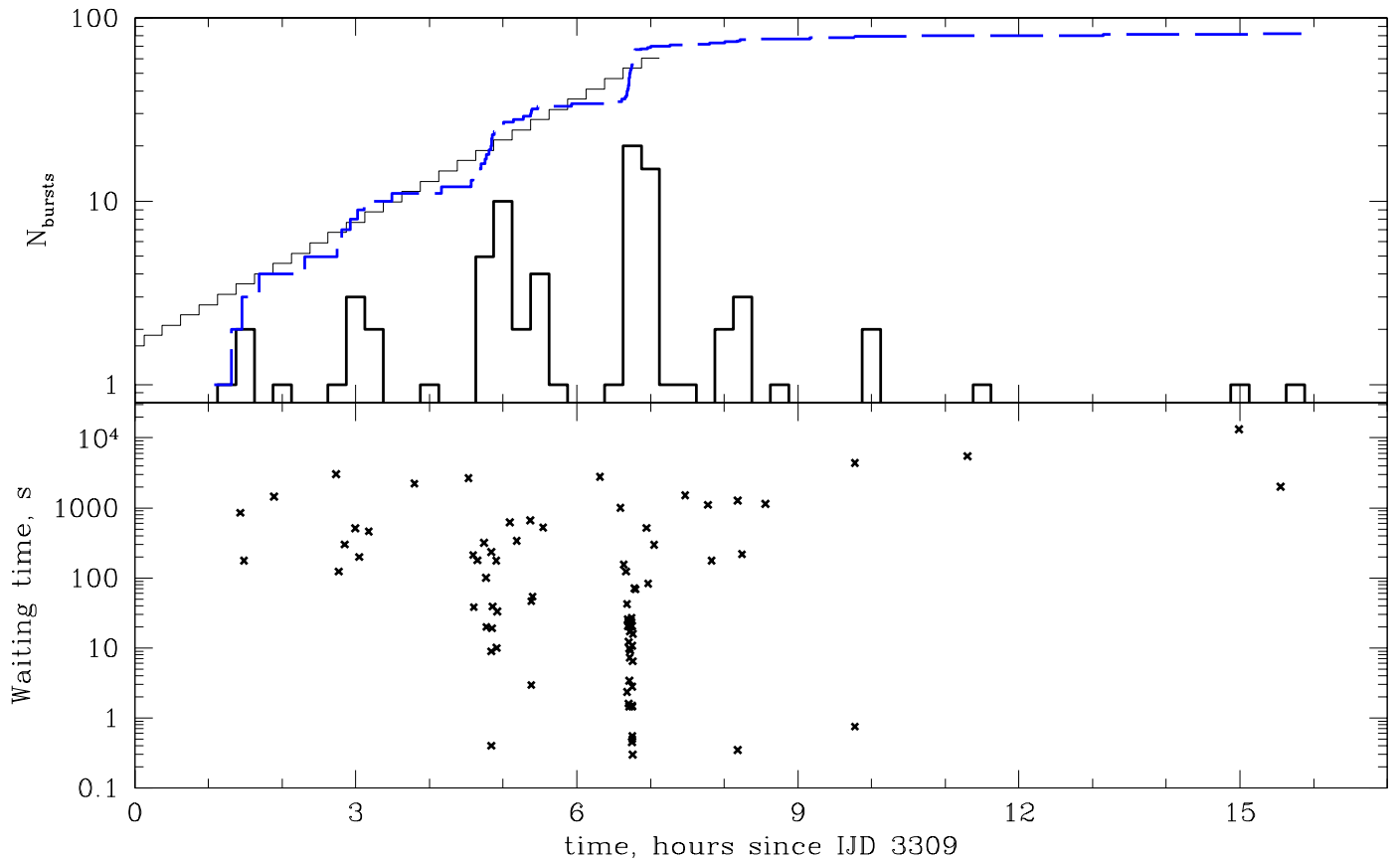}
\caption{{\it Top}: the thick solid line is the rate of bursts detected both in IBIS/ISGRI and SPI/ACS as a function of time on 22 January 2009. The dashed blue histogram shows the cumulative burst distribution.The thin straight solid historgam shows a fit with an exponentially rising burst rate with rise time $T_r=1.94$~hr. The dotted thin solid histogram shows a fit with the normal distribution with the width $\sigma_T= 2.64$~hr. {\it Bottom}: the waiting time between subsequent bursts as a function of time.}
\label{fig:overall_lc}
\end{figure*}

\axp\ was discovered by the {\it Einstein} satellite \citep{lamb81} in the search of an X-ray counterpart of a {\it COS-B} source and was recently identified as a magnetar in the center of the supernova remnant candidate SNR G327.24-0.13 \citep{gelfand07}. A discovery of pulsed radio emission with a period $P\simeq 2$~s and period derivative $\dot P\simeq 2.3\times 10^{-11}$~s$^{-1}$ enabled the estimation of the magnetic field close to the neutron star $B\simeq 2.2\times 10^{14}$~G as well as of its distance $D\simeq 9$~kpc \citep{camilo07}. The X-ray flux from the source is known to exhibit large variations in the range of $(0.1- 5)\times 10^{-12}$~erg~cm$^{-2}$s$^{-1}$ in the 1-8~keV energy band, which corresponds to variations of the source luminosity $10^{34}$~erg~s$^{-1}$ $\le L_{\rm X}\le 10^{35}$~erg~s$^{-1}$ \citep{halpern08}. The source spectrum in the soft X-rays is dominated by a quasi-thermal emission with the temperature $T\simeq 0.4 - 0.5$~keV. \citet{halpern08} reported the discovery of X-ray pulsations from the source with the period being consistent with that of the radio pulsations. 

In this paper we report on a study of the episode of bursting activity of \axp\ in January 2009. During this episode, about $200$ bursts from the source \citep{atel1,mereghetti09} were detected by the Anti-Coincidence Shield (ACS) of the spectrometer SPI on board the {\it INTEGRAL}  satellite \citep{winkler03}. The flaring activity of the source was initially discovered by {\it Swift} \citep{gcn_swift} and was also observed by the {\it Fermi/GBM} telescope \citep{gcn_fermi,gcn_fermi2}. The source was not detected in the radio band during the peak of its high-energy activity \citep{atel_radio_no}. Pulsed radio emission appeared about three days after the major activity episode \citep{atel_radio}. 

Major episodes of activity consisting of clusters of $\sim 10-10^2$ bursts are typical for SGRs, but in the case of AXPs such activity was detected only once, in the source \axpp\ in 2003 \citep{gavriil04}. Below we compare the characteristics of the bursting activity of \axp\ with those of \axpp\ and of SGRs.  We find that contrary to the \axpp\ bursts, the luminosity of the \axp\ bursts at the energies above $\sim 80$~keV during the January 2009 activity period was comparable to the one of  SGR bursts and possibly reaching the luminosity scale of  SGR giant flares. 

Detection of the increased activity of the source on 22 January 2009 has led to a dedicated target-of-opportunity (TOO) observation of the source during the period 24-29 January 2009. During the TOO on-axis observation, the rate of bright bursts as well as the luminosity of the individual bursts largely decreased. However, the source was detected at an average flux level $(1.9 - 2.6)\times 10^{-10}$~erg~cm$^{-2}$~s$^{-1}$ in the 20-40~keV band, which is still two to three orders of magnitude higher than the "quiescent" and "flaring" flux previously measured by {\it XMM-Newton} at somewhat lower energies \citep{baldovin09,denhartog09}. In spite of this decrease, the brightness of the bursts was enough to saturate the telemetry of the imager IBIS/ISGRI on board of \intgr, because during the TOO observation the signal from the source was not suppressed by the walls of the telescope. The analysis of the ISGRI detector lightcurves and of the data of the anti-coincidence shield (ACS) of the spectrometer SPI during the 22 January 2009 activity period is still the best suited for the study of the spectral and timing characteristics of the individual bursts/flares of the source as well as for the statistical study of the properties of the bursts during this exceptional activity period. Below we concentrate on this analysis.  

\section{Data analysis}

The ACS consists of 91 bismuth germanate (BGO) scintillator crystals of a thickness  of between 16 and 50 mm and with a total mass of 512 kg \citep{vonkleinin03}. The ACS events are recorded as the overall detector count rate sampled in time intervals of 50 ms. No energy or directional information is available. The ACS is sensitive to photons above a low-energy threshold corresponding to approximately 80 keV. Dead time and saturation effects in the crystals and electronics are negligible ($<1$\%) for total count rates smaller than a few times $10^5$~cts~s$^{-1}$, but become essential for rates as high as $10^6$~cts~s$^{-1}$ \citep{mereghetti05}, which is the case for the brightest bursts from \axp, reported in our paper.

The ISGRI detector \citep{lebrun03}  is part of the IBIS telescope on-board of {\it INTEGRAL}. It is an array of $128\times 128$ pixels made of semiconductive CdTe, sensitive to photons between 15 keV and 1 MeV. ISGRI works in photon-by-photon mode, so that lightcurves with time bins as short as $60\ \mu$s can be produced. The walls of the collimator above the  ISGRI detector are made of lead and act as a shield to photons with energies of up to 200 keV. For geometrical reasons, the optical depth of the shield is smaller for photons arriving at large off-axis angles. Hard photons from off-axis sources can pass through the shield and reach the detector, so that bright sources, such as $\gamma$-ray bursts, can be detected also when they are formally outside the field of view (see e.g. \citet{marcinkowski06}). 

For our analysis we have extracted the ACS and ISGRI detector lightcurves from the data of {\it INTEGRAL} revolution 766, which covers a period from 2009-01-20T14:34 to 2009-01-23T04:23 UTC. We have used the Offline Science Analysis (OSA v. 7.0) package, provided by the ISDC \citep{courvoisier03} to reduce the data.

During revolution 766 the {\it INTEGRAL} instruments were pointing toward the 3C 279 region, and  \axp\ was at $\sim 60^\circ$ off-axis angle, outside the field of view (FoV) of the {\it INTEGRAL} instruments. Using the {\tt ii\_light} tool we have extracted ISGRI detector lightcurves in two energy bands, $20-60$~keV and $60-200$~keV. We have applied the barycentric time correction using the  {\tt barycent}\footnote{{\tt barycent} tool is a part of the {\it INTEGRAL} Offline Science Analysis (OSA) package distributed by ISDC \citep{courvoisier03}. tool and substituting the source coordinates found by  \citet{camilo07}}. To perform a comparison between the signals in ISGRI and ACS, we have binned the ISGRI lightcurve into 50-ms long time bins in order to match the bins of the ACS lightcurve.

To identify the individual bursts, we have used the ACS lightcurve,
which had significantly higher signal statistics. The bursts were
identified using the following algorithm. We have measured  the background count rate
via calculation of 10-seconds running mean of the ACS lightcurve. Next, we have identified the moments of the on-set of individul bursts as the moments when the count rate rises beyond the $5\sigma$ level above the background.  The burst duration  was determined as the time interval between the moments when the count rate rises beyond the $5\sigma$ level and the moment when the count rate drops below the $3\sigma$ level above the background. The error on the burst duration was calculated as the difference between the maximum and minimum estimates of the
burst duration, which were defined as the time intervals between the moments when the lightcurve crosses 6 and 4 $\sigma$ above the background rate for the maximal duration and 4 and 2 $\sigma$ for the minimum duration, respectively. Additionally we have taken into account the 50 ms systematic error determined by the fixed size of the ACS time bin.

The ACS signal is known to occasionally contain short bursts of unidentified nature (possibly due to particle precipitations or due to electronics noise) of a duration of 50-100~ms. Such bursts could be confused with the real bursts from \axp. To separate the \axp\ bursts from the possible instrumental effects, we  have searched for each of the identified bursts for a counterpart of the burst in the ISGRI ligthcurve. Only the ACS bursts found simultaneously  in ISGRI detector lightcurves with a significance higher than 3 sigma were accepted for the statistical studies. This selects 84 out of about $200$ bursts detected in the ACS lightcurve during the analyzed period. Taking into account the smaller collection area of ISGRI, such a strict selection criterion rejects a large number of weaker bursts from \axp. However, it assures that all the burst data are "instrumental background free", which provides an advantage from the viewpoint of statistical studies and also selects the bursts for which information about the energies of the detected photons is available.

\section{Results}

\subsection{The 22 January 2009 activity episode} 

Figure~\ref{fig:overall_lc} shows the evolution of the source during the bursting activity period. The upper panel of the figure shows the evolution of the rate of bursts on 22 January 2009. The maximum of burst activity happened around UTC 22-01-09T06:40 with more than 40 bursts detected within half-an-hour. 
This major activity episode was preceded by several weaker episodes, with a growing peak burst rate. A fit of the evolution of the burst rate with an exponential function, $N\propto \exp\left[(t-t_{\rm max})/T_r\right]$, gives the rise time $T_r=1.94\pm 0.14$~hr. A fit with normal distribution, $N\propto \exp(-(t-t_{\rm max})^2/(2\sigma_T^2)$ gives a comparable rise time, $\sigma_T=2.64\pm 0.05$~hr, (see upper panel of Fig. \ref{fig:overall_lc}). The increase of the burst rate ended abruptly half an hour after the major activity episode. The burst rate has subsequently continued to decrease on several hour- and day-time scales. It is worth noting that the source continues to produce bursts occasionally, but with a much lower rate one month after the peak of activity \citep{atel_february,atel_february1}.

The lower panel of Fig.~\ref{fig:overall_lc} shows the evolution of the waiting time between the subsequent bursts during the activity period. One can see that the exponential rise in the burst rate during the precursors of the major activity episode was accompanied by an exponential decrease of the minimal waiting time between the subsequent bursts, down to $\sim 0.2$~s (which is much shorter than the inverse of the burst rate). After the end of the major episode, the waiting time has increased abruptly  to $\ge 100$~s and has continued to steadily increase on a time scale of several hours.

An expanded view of the ACS lightcurve during the major outburst is shown in Fig.~\ref{fig:major_zoom}. The brightest event during the activity period was the burst marked {\bf b} in this figure, with the ACS count rate reaching $1.8\times 10^6$~cts~s$^{-1}$ \citep{atel1}. Several bursts of comparable strength were observed within an interval of about $2$~hours around the strongest event. 

\subsection{Individual burst lightcurves}

\begin{figure}
\includegraphics[width=\linewidth]{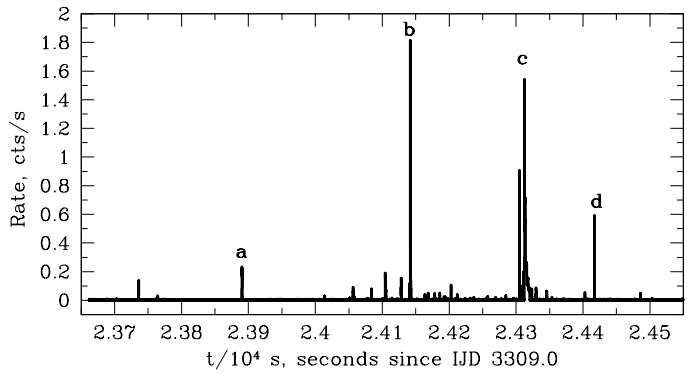}
\caption{SPI/ACS lightcurve during the main activity episode of \axp\ between 6:30 and 7:00, 22 January 2009 (UTC).}
\label{fig:major_zoom}
\end{figure}

\begin{figure*}
\includegraphics[width=\linewidth]{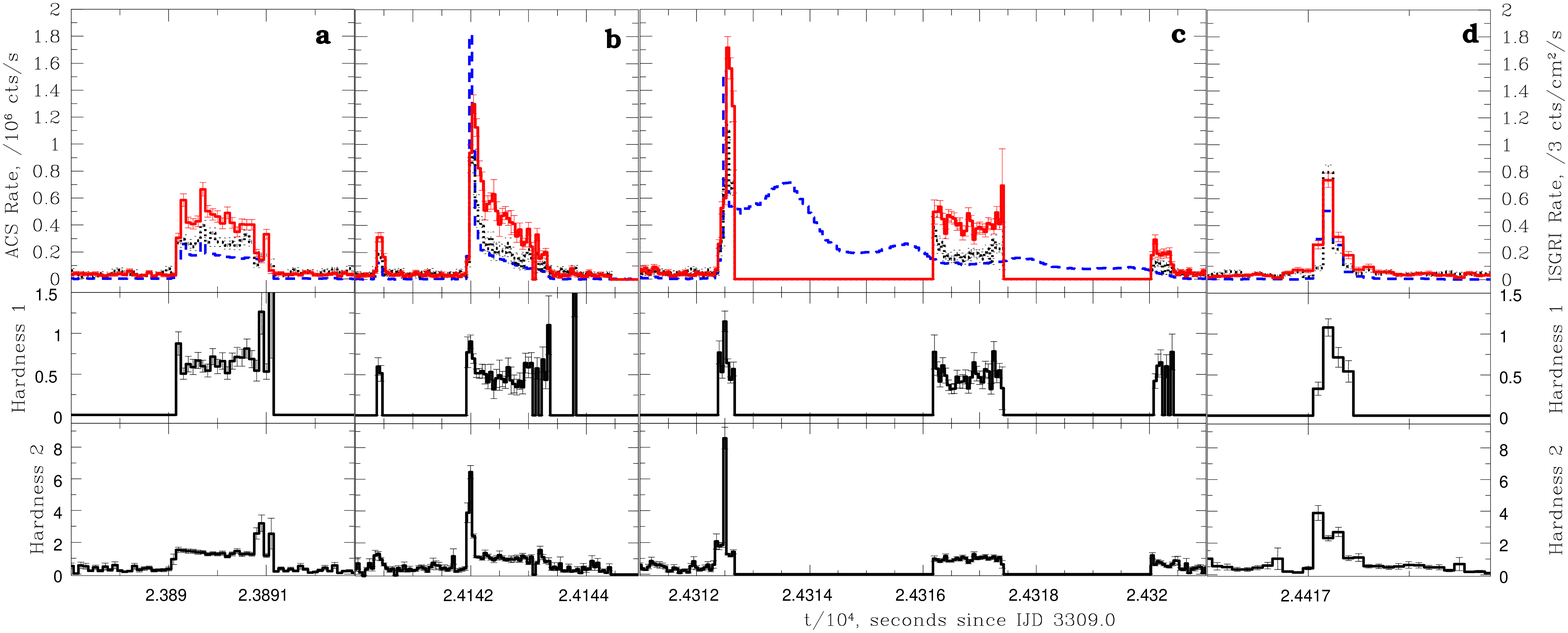}
\caption{The lightcurves of the bursts marked as {\bf a,b,c} and {\bf d} in Fig.~\ref{fig:major_zoom}. Red solid (black dotted) line shows the ISGRI detector lightcurve in the 20-60~keV (60-200 keV) energy band, binned in 50~ms time bins. The periods of zero count rate correspond to the gaps which occurred because of the telemetry saturation. Blue dashed blue line shows the SPI/ACS lightcurve. Signal oscillations visible in the long burst {\bf c} are with the period $\simeq 2$~s of rotation of the neutron star.  For each burst, the middle panels show the evolution of the hardness ratio between the 60-200 keV and 20-60~keV flux in ISGRI. The lower panels show the evolution of the ratio of the SPI/ACS flux to  the 20-60~keV flux in ISGRI. }
\label{fig:burst_lcs}
\end{figure*}

An expanded view of the strongest bursts, marked {\bf a}, {\bf b}, {\bf c} and {\bf d} in Fig.~\ref{fig:major_zoom} is shown in Fig.~\ref{fig:burst_lcs}. The upper panels of the figure show the evolution of the SPI/ACS ($R_{\rm ACS}$, measured in cts~s$^{-1}$) and ISGRI 20-60~keV ($R_{20-60}$, measured in cts~cm$^{-2}$s$^{-1}$) and 60-200~keV ($R_{60-200}$, measured in cts~cm$^{-2}$s$^{-1}$) count rates. The middle panel shows the evolution of the ISGRI-ISGRI hardness ratio $H_1=R_{60-200}/R_{20-60}$, while the lower panels shows the evolution of the ACS-ISGRI hardness ratio, defined as $H=10^{-6}R_{\rm ACS}/R_{\rm 20-60}$.

Remarkably, all the four strongest bursts which happened during the main activity episode have qualitatively different time profiles. The lightcurve of  burst {\bf a} has a "flat top" shape with a sharp rise and decay and a nearly constant flux "plateau" period of $\sim 1$~s duration. This shape is similar to the one of the "precursor" to the bright flare of the SGR 1806-20 reported by \citet{nature05}. During the "plateau" phase the spectrum of the source does not vary, as one can see from the lower panel of Fig.~\ref{fig:burst_lcs}a. However, the width of the plateau seems to increase with decreasing photon energy, as one can see from the evolution of the ISGRI-ISGRI and ACS-ISGRI hardness ratios.

The strongest burst shown in Fig.~\ref{fig:burst_lcs}b consists of a bright short spike of $\sim 100$~ms  duration, and a $\sim 1$~s long softer "tail". The peak count rate in the ACS detector is $\sim 1.8\times 10^6$~cts~s$^{-1}$. This count rate is, in fact, close to the peak count rate observed in the giant flare of SGR 1820-06 by \citet{mereghetti05}. However, this does not necessarily mean that the peak fluxes of the two events were comparable. The problem is that at such a high count rate the ACS signal is known to be affected by saturation effects, so that, independently of the real $\gamma$-ray flux, the count rate in the ACS detector is close to the maximum possible one. 

The uncertainty of the energy response of the ACS detector does not allow a direct conversion of the observed count rate into physical flux units. An order-of-magnitude estimate of the peak flux could be obtained in the following way. For the moderate off-axis angles, the effective area of the ACS detector rises from $A_{\rm eff,ACS}(0.2\mbox{ MeV})\sim 3\times 10^2$~cm$^2$ up to $A_{\rm eff,ACS}(1\mbox{ MeV})\simeq 3\times 10^3$~cm$^2$ in the energy range of 0.1-1~MeV \citep{mereghetti05} (we use these numbers for an order-of-magnitude estimate, so that a factor of $\sim 1$ difference in $A_{\rm eff,ACS}$ for the off-axis angles in the range $\theta\sim 45^\circ$ can be neglected). The average  energy threshold of the detectors is $E_{\rm thr}\simeq 80$~keV \citep{vonkleinin03}. To estimate the energy flux from below one can assume that all the detected photons have an energy about the low-energy threshold, $E_\gamma\sim 0.1$~MeV. Multiplying this energy by $R_{ACS}$ and dividing by the effective area one finds a lower limit on the flux\footnote{the analysis of \citet{mereghetti09}, which appeared after this paper was submitted, results in a comparable estimate of the lower limit on the source flux.} 
\begin{eqnarray}
&&F\ge \frac{E_\gamma R_{\rm ACS}}{A_{\rm eff,ACS}}\simeq 3\times 10^{-4}\left[\frac{E_\gamma}{0.1\mbox{ MeV}}\right]\nonumber\\ && \left[\frac{R_{\rm ACS}}{2\times 10^6\mbox{ cts~s}^{-1}}\right]\left[\frac{A_{\rm eff,ACS}}{10^3\mbox{ cm}}\right]^{-1} \mbox{erg~cm}^{-2}\mbox{s}^{-1}.
\end{eqnarray}
At the distance $D\simeq 9$~kpc this implies the luminosity
\begin{equation}
L_{\rm\bf b}= 4\pi D^2F\ge 3\times 10^{42}\mbox{erg~s}^{-1}.
\end{equation} 
This luminosity is  several orders of magnitude larger than that of the bursts of the AXPs observed up to now (see Table \ref{tab:bursts}).  At the same time, the luminosity of the bursts of \axp\ turns out to be comparable to the one of a typical SGR burst (see e.g. \citealt{mereghetti08}).

The short initial spike of burst {\bf b}  is less pronounced in the ISGRI lightcurves. This might be due to the fact that in spite of the significant absorption of the signal from the source by the walls of the IBIS telescope, the signal in the ISGRI detector is strongly affected by saturation effects, while in the ACS detector the saturation is less pronounced. Otherwise, the difference in the strength of the spike in the ISGRI and ACS lightcurves could be due to the hardness of the spectrum of the spike. 
A "hard spike plus soft extended tail" morphology is typical for the SGR bursts  \citep{gogus00} (see, however, Ref. \citet{ibrahim01} who report a powerful SGR burst with a pulsating tail but without hard onset  spike).

The two possibilities can be, in principle, distinguished via an analysis of the hardness ratio in the ISGRI detector alone. Indeed, the soft (20-60~keV) and hard (60-200~keV) band fluxes in ISGRI should be affected by  saturation effects in a similar way, so that the hardness ratio does not change. The middle panels of Fig.~\ref{fig:burst_lcs} show the evolution of the ISGRI hardness ratio over the four brightest bursts. One can see that the spikes of the bursts {\bf b} and {\bf c} indeed have somewhat harder spectra than the afterglows. However, the jump in the ACS/ISGRI hardness ratio during the spike is more pronounced than in the ISGRI/ISGRI hardness ratio. This indicates that both the above mentioned effects might be present:  the spectrum of the spike is harder than the one of the afterglow and  the signal in ISGRI is much stronger affected by the saturation effects than that in the ACS. 

Not only the bright spike, but also the afterglow lightcurve of  burst {\bf b} in the ISGRI detector suffers from a saturation effect, which is visible as an anomaly in the distribution of waiting times between subsequent photons. This saturation effect results in the production of short time gaps of $\sim 16, 32$ or $64$~ms duration, which appear in a quasi-periodic manner. We have corrected the ISGRI lightcurve shown in Fig.~\ref{fig:burst_lcs}b for this effect by excluding the gaps from the analysis. We have checked that this photon counting problem affects only the brightest burst and that no anomalies of the waiting times between subsequent photons are present in the spikes and afterglows of the other bursts.

The 100~ms burst phase of the brightest event is followed by a softer "afterglow" tail which is characterized by  a power-law decay profile $F(t)\sim F_0/(1+[t/T_{\rm dec}])$
with $T_{\rm dec}\simeq 10$~ms for the ACS and ISGRI 60-200~keV energy band and $T_{\rm dec}\simeq 30$~ms in the ISGRI 20-60~keV energy band. The powerlaw decay ends abruptly $\sim 1$~s after the start of the burst, which might indicate a more complicated behavior of the afterglow emission, e.g. the presence of pulsations (see below).

\begin{table}
\begin{tabular}{llll}
\hline
Name & $F$& $L$& Reference\\ & [$10^{-8}$~erg~cm$^{-2}$s$^{-1}$] & [erg~s$^{-1}$] & \\
\hline
4U 0142+61 & $0.3\div 4$~(2-30 keV)&$\le 1.7\times 10^{40}$&1\\
 1E 1048-59 &  $0.5\div 1.1$~(2-40 keV)&$\le 1.6\times 10^{37}$&2\\
\axp& $>3\times 10^4$~($>80$ keV)&$>2.7\times 10^{42}$&this work\\
CXOU J1647-45 & $\simeq 1$~(15-150 keV)&$\le 1.7\times 10^{37}$&4\\
XTE J1810-197 & $0.07\div 10$~(2-30 keV)&$\le 10^{38}$&5\\
1E 2259+586 & $0.1\div 40$~(2-60 keV)&$\le 2.5\times 10^{39}$&6\\ 
\hline
\end{tabular}
\caption{Comparison of peak fluxes and luminosities  of the AXP bursts. References: 1-\citet{gavriil07}; 2-\citet{gavriil02}; 3-this work; 4-\citet{israel07}; \citet{woods05};  6-\citet{kaspi03,gavriil04}.}
\label{tab:bursts}
\end{table}

A somewhat weaker burst shown in Fig.~\ref{fig:burst_lcs}c exhibits a spike plus afterglow morphology similar to the one of burst {\bf b}, but with a much larger fraction of the power emitted during the afterglow phase. Two gaps in the ISGRI detector light curve appear because of the telemetry saturation. This prevents an analysis of the spectral evolution during the afterglow phase, but, as one can see from the lower panel of Fig.~\ref{fig:burst_lcs}c, the ISGRI count rate closely follows the ACS count rate, so that the hardness ratio does not change over the time interval where ISGRI data are available. In the ACS lightcurve of the burst, shown by the blue dashed line, one can clearly identify the oscillations with the period $\simeq 2$~s, close to the period of rotation of the neutron star. These oscillations were previously reported by \citet{atel2}.

The burst {\bf d} shown in the Fig.~\ref{fig:burst_lcs}d consists of a single hard spike of the duration of $\sim 100$~ms, with almost no detectable afterglow. The ratio of the peak luminosity of the spike to the peak luminosity of the afterglow in this burst  and the hardness of the spike emission are comparable to the one of the strongest burst {\bf b}. 

\subsection{Statistical study of the burst properties}

\begin{figure}
\includegraphics[width=\linewidth]{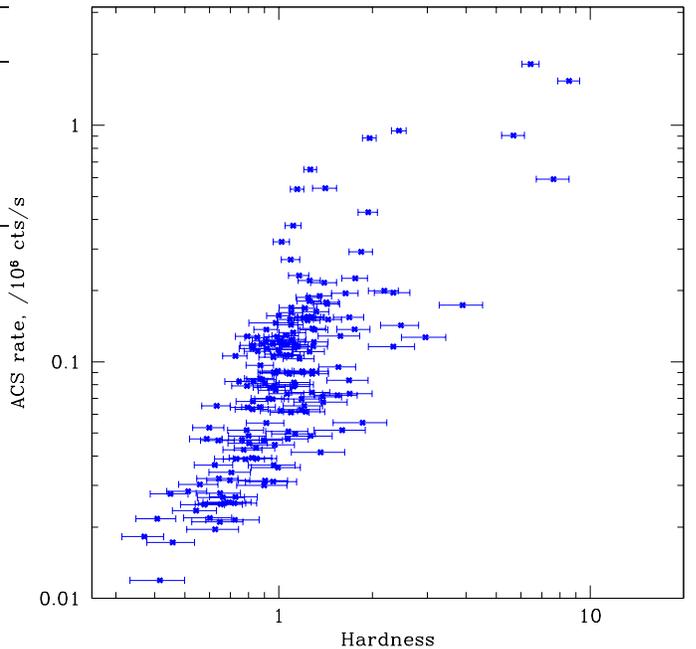}
\caption{Hardness as a function of ACS count rate during the main activity period of the source shown in Fig.~\ref{fig:major_zoom}.}
\label{fig:hardness_acs}
\end{figure}

From Fig.~\ref{fig:burst_lcs} it is clear that the bright short spikes in the burst lightcurves are characterized by comparable ACS/ISGRI hardness ratios. It is not clear a priori if all the short spikes have comparably hard spectra, independent of their strength, or whether the hardness ratio of the spike/afterglow emission is determined by the burst luminosity. In order to distinguish between these two possibilities, we plot in Fig.~\ref{fig:hardness_acs} the hardness ratio as a function of the ACS count rate during the main activity period.  From this figure one can see a correlation between the hardness and the count rate, which indicates that brighter bursts and afterglows also tend to have harder spectra. The value of the hardness of the brightest bursts is affected by the saturation effects in ISGRI and possibly also in the ACS detectors, but for the moderate flux values the observed  flux-hardness correlation is robust. The dependence of the hardness of the spectrum on the luminosity was noticed in the bursts of \axpp\ by \citet{gavriil04}. This type of dependence is the opposite of those observed in SGR bursts, for which a softening of the spectrum with increasing luminosity was observed \citep{gogus00}. 

\begin{figure}
\includegraphics[width=\linewidth]{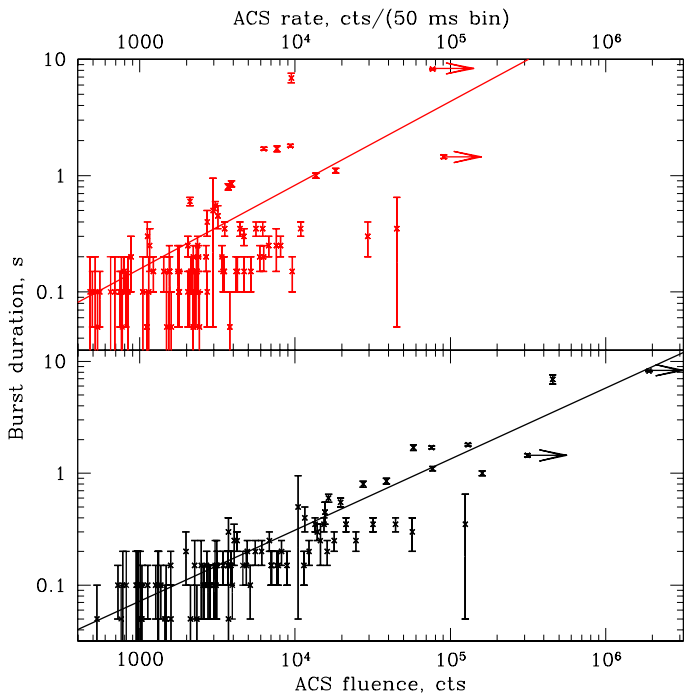}
\caption{Burst duration as a function of the ACS peak rate (top panel) and fluence (botttom panel). Solid lines show the powerlaw fits to the data points. The error of the burst duration is assumed to be equal to the size of the ACS lightcurve time bin, $50$~ms. Points shown as lower limits on the flux/fluence are those affected by possible saturation effects in ACS detector.}
\label{fig:flux_duration}
\end{figure}

Figure~\ref{fig:flux_duration} shows the dependence of the burst duration on the peak count rate and overall fluence of the bursts in ACS. The burst duration is defined as the time interval between the start and end of the burst, defined as explained in Sect. 2. In our case this duration measure is equivalent to the so-called "$T_{90}$" (the duration of time interval in which 90\% of the burst fluence is contained). This is explained by the fact that signal statistics in ACS are very high for all the bursts retained for the statistical analysis.  It is clear that if a significant fraction of the burst fluence is emitted in the extended afterglow, a correlation between the burst duration and the fluence is expected, because the afterglows of stronger bursts remain detectable during longer periods of time. On the other hand, if most of the burst fluence is emitted in a short spike, no correlation between the burst duration and the fluence would be present. Figure~\ref{fig:flux_duration} indicates that in most of the bursts a significant contribution to the total energy output comes from the afterglow emission. On the other hand, one can see that the burst duration $T$ is correlated with the peak flux, i.e. bursts with stronger spikes possess on average stronger afterglows. If the correlation of the burst durations with burst fluxes and fluences are fit by power law, the best-fit powerlaw dependences are $R\sim T_{\rm peak}^{1.4 \pm 0.2}$ and \textbf{$F\sim T^{1.58\pm 0.08}$} for the peak flux and fluence respectively. It is clear from Fig.~\ref{fig:hardness_acs} that the spread of the data points around the powerlaw model fit is quite large, which explains the very high reduced $\chi^2_\nu$ values of the fit of the peak flux vs. duration correlation, $\chi^2_\nu=19.6$ (for 82 d.o.f.). The fluence vs. duration correlation is somewhat tighter and the quality of the powerlaw fit is somewhat better,  $\chi^2_\nu=9.3$ (for 82 d.o.f.). The observed fluence-duration correlation is different from the one observed in \axpp\ ($F\sim T^{0.54\pm 0.08}$ by \citet{gavriil04}) and instead resembles more the fluence-duration relation of SGR bursts, e.g. $F\sim T^{1.13}$ in \mbox{SGR 1900+14}, \citep{gogus99} (note that the correlation $F-T$ correlation is not very tight both in our case and in the case of SGR 1900+14, so that the two relations are consistent).

\begin{figure}
\includegraphics[width=\linewidth]{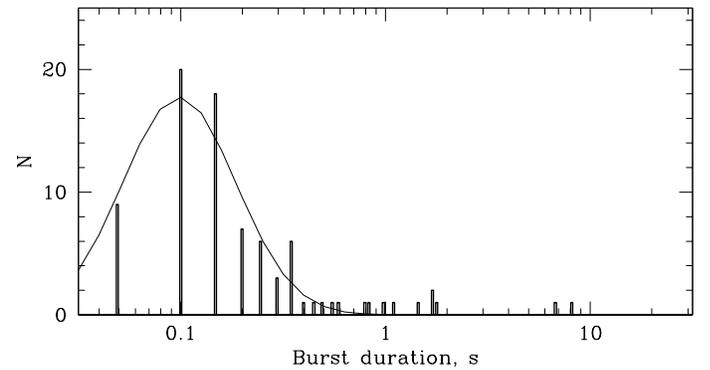}
\caption{Distribution of burst durations in ACS. Thin solid curve shows the best fit log normal distribution.}
\label{fig:duration_histogram}
\end{figure}

From Fig.~\ref{fig:burst_lcs} it is clear that already in the forth-brightest burst {\bf d} the afterglow emission is difficult to detect because of the much lower signal statistics. This means that for most of the weaker bursts, visible in Fig.~\ref{fig:major_zoom}, only the bright spike is detectable. The duration of the spikes in the bursts {\bf b, c} and {\bf d} shown in Fig.~\ref{fig:burst_lcs} is 1-2 ACS time bins, i.e. $\le 100$~ms. One expects that the $\sim 100$~ms spike duration should give a correct estimate of the typical durations of the majority of the weaker bursts detected by the ACS during the activity of the source. This is confirmed by Fig.~\ref{fig:duration_histogram} in which the distribution of durations of the bursts detected simultaneously by the ACS and ISGRI detectors is shown, peaking at $100$~ms. The $50$~ms time resolution of the ACS detector does not allow the measurement of  details of the burst duration distribution on shorter time scales. Fitting the distribution of burst durations with a log-normal distribution we find a mean of $T=68\pm 3$~ms and a scatter of $30$~ms$<T<155$~ms. The excess of longer duration bursts in the $0.5-10$~s interval is due to the presence of the extended afterglows and of several "plateau" like events, similar to burst {\bf a} shown in Fig.~\ref{fig:burst_lcs}a.

\begin{figure}
\includegraphics[width=\linewidth]{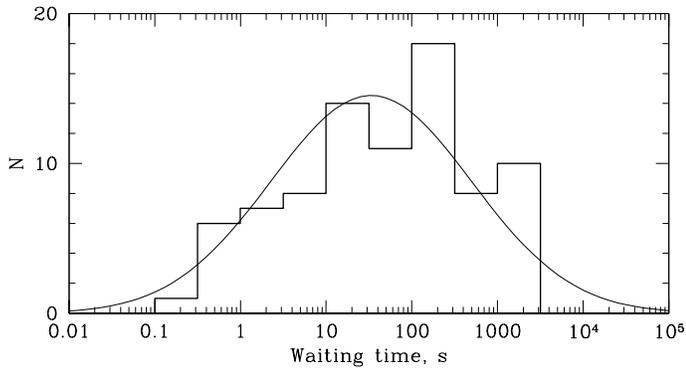}
\caption{Distribution of the waiting times between subsequent bursts.}
\label{fig:wt_distribution}
\end{figure}

From the bottom panel of Fig.~\ref{fig:overall_lc} it is clear that the main activity period of the source is characterized by the extremely short intervals between subsequent bursts. The waiting time decreases down to  values comparable to the burst durations at the peak of activity. Figure~\ref{fig:wt_distribution} shows the distribution of waiting times between the bursts for the entire activity period on 22 January 2009. The waiting time distribution can be fit by a wide log-normal distribution with a mean $T_{\rm waiting}=34\pm 11$~s and a scatter of $2.3~s <T_{\rm waiting}<496$~s. At the peak of activity, the waiting time between subsequent bursts decreases to a time scale comparable to the duration of individual spikes, so that the bursts can be considered as either separate events or as parts of the "multi-spike" flares. This behavior is similar to the one observed in SGR bursts \citep{gogus99}. At the same time, the observed scatter of the waiting times is larger than the one found in the {\it RXTE} observations of \axpp\ by \citet{gavriil04}, who found that the waiting time between subsequent bursts is always larger than $1$~s ($\ge 10$ times longer than the typical burst duration). The difference between the minimal waiting times between subsequent bursts in \axp, reported here, and that of the bursts of \axpp, reported by \citet{gavriil04}, could have four different explanations:
\begin{enumerate}
\item the bursts of 1E 2259+586 and of \axp\ are physically different;
\item the {\it RXTE} observations of 1E 2259+586 missed the peak of the bursting activity of the source;
\item a gap between the minimal waiting time and burst duration in the 1E 2259+586 observations is related to the sensitivity limit of  {\it RXTE} (weaker bursts remained undetected);
\item the definition of "burst event" used by \citet{gavriil04} slightly differs from the one used in our analysis.\footnote{ E.g. two individual bursts with short waiting time $T_{\rm waiting}\ll 1$~s might have been considered as a single multi-peaked burst in the analysis of \citet{gavriil04}.}
\end{enumerate}
The most natural explanation seems to be the second one:  during the limited time of the {\it RXTE} observations an intermediate state of activity of \axpp\ was caught. This is clear from the fact that the waiting time between the subsequent bursts in the {\it RXTE} observations has systematically evolved toward higher values during the observation. Such an evolution differs from the one found in the case of \axp\ (bottom panel of Fig.~\ref{fig:overall_lc}), where the waiting time first decreases to the lower values as the source activity increases.

The arrival times of bursts of AXPs are found to be correlated with the pulsed emission phase. In the case of \axp\ this fact is difficult to verify, because the amplitude of the pulsed quiescent X-ray emission is just some $7$\% of the quiescent X-ray flux \citep{halpern08}.  It is clear from  Fig.~\ref{fig:burst_lcs}c that the amplitude of the pulsed emission significantly increases during the afterglows of the bright bursts. To test if the arrival times of the individual bursts are correlated with the phase of the pulsed radio/X-ray emission, one can 
\begin{itemize}
\item study the distribution of the phases of arrival of the short spikes and/or
\item compare the phases of the maxima of the pulsed tails of the brightest bursts.
\end{itemize}

\begin{figure}
\includegraphics[width=\linewidth]{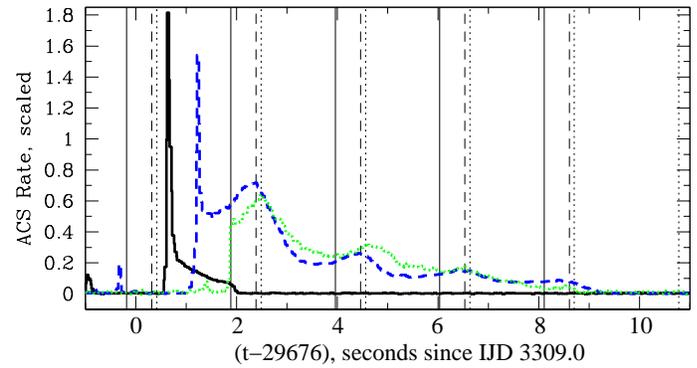}
\caption{Relative phases of the pulses of the burst afterglows. The green dotted line shows a bust which occurred on January 22, 8:18 am. Black solid line shows the brightest burst {\bf b} shifted by 2671 periods forward in time. The blue dashed line shows the second-brightest burst {\bf c} shifted by 2589 periods.  The count rate of this burst is multiplied by a factor of three to highlight the similarity with the time profile of the burst {\bf c}. The vertical solid and dashed line shows the phases of the maxima of the pulses of the burst afterglows. The vertical solid lines are shifted by $\Delta\phi=0.25$ with respect to the vertical dotted lines.}
\label{fig:burst_oscillations}
\end{figure}

Figure~\ref{fig:burst_oscillations} shows a comparison of the afterglows of three bright bursts. The burst shown by the solid line is the brightest burst {\bf b}, the burst shown by the dashed line is the second brightest burst {\bf c} and the burst shown by the dotted line is a somewhat weaker burst, which happened two hours after the main activity episode. To shift the lightcurves of the individual bursts by an integer number of periods, we use the measurement of the period of \axp\ reported by \citet{atel_period}. From this figure one can notice that although the phases of the maxima of the pulsed afterglow do not coincide exactly, they are rather close to each other. If the afterglow is due to emission from the surface of the neutron star, this might indicate that the afterglows of different bursts are produced by one and the same "hot spot" on the neutron star's surface or that the "trapped fireball" in the neutron star's magnetosphere always appears at the same latitude / longitude. For comparison, we show in the same figure the ACS lightcurve of the brightest burst {\bf b}. The phase of the sharp end of the afterglow of burst {\bf b} is shifted by $\delta\phi\sim 0.25$ with respect to the phases of the maxima of the pulsed afterglow emission of the other bursts. If the sharpness of the termination of the afterglow of burst {\bf b} is ascribed to the presence of pulsations in the afterglow emission,  the location of the afterglow-emitting region of this burst should be displaced. Otherwise, the difference in the strength and shift in the phase between the afterglows of the bursts {\bf b} and {\bf c} could be due to different physical mechanisms of the afterglow production.   

\begin{figure}
\includegraphics[width=\linewidth]{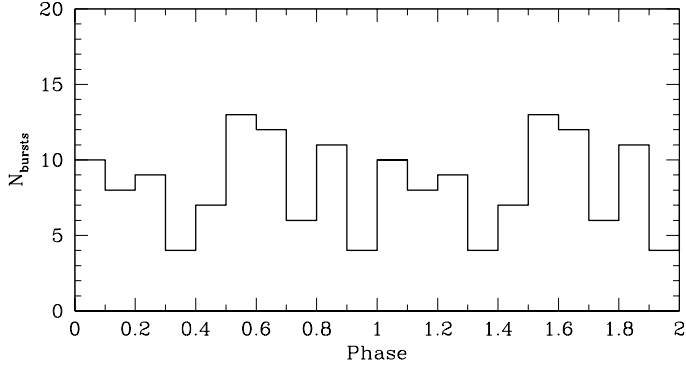}
\caption{Distribution of the burst start times in terms of the phase of the pulsed emission. }
\label{fig:bursts_phase}
\end{figure}

From Fig.~\ref{fig:burst_oscillations} one can notice that, contrary to the pulsed afterglows, the short spikes of different bursts do not arrive in phase. This is confirmed by Fig.~\ref{fig:bursts_phase}, which shows the distribution of the number of bursts as a function of the phase of the pulsed emission. The phase $\phi=0$ is assumed to be the phase of the sharp end of the afterglow of burst {\bf b}. One can see that the phases of the bright spikes are randomly distributed over the pulse period. 

It is worth noting that the short spikes indeed seem to arrive at random times and that properties (flux, arrival time) of the spike are not correlated with the properties of the afterglows. For example, the burst shown by the green dotted line in Fig.~\ref{fig:burst_oscillations} has a sharp start of the afterglow, but does not possess an initial spike at all. We have also found bursts in which the moment of arrival of the short spike is delayed compared to the moment of the onset of the "afterglow-like" emission. 

\begin{figure}
\includegraphics[width=\linewidth]{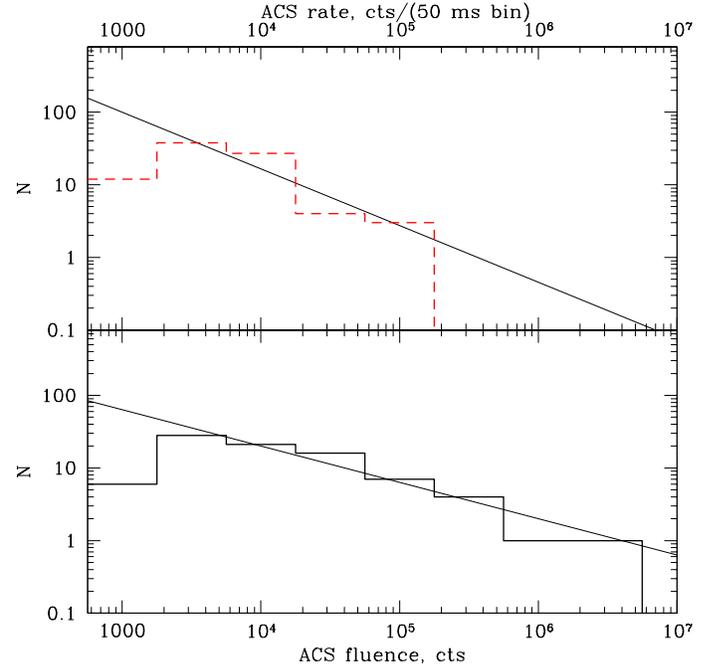}
\caption{Top panel: distribution of the ACS peak count rates of the bursts detected by both ACS and ISGRI detectors. Bottom panel: distribution of fluences of the bursts. The thin straight lines show powerlaw fits to the distributions.}
\label{fig:fluence}
\end{figure}

The large effective area of the ACS detector enables us to detect the bursts with a fluence down to three orders of magnitude lower than the one of the brightest bursts.  The distribution of the burst fluences is shown in Fig.~\ref{fig:fluence}. The fluence distribution follows a power law with an exponent $-0.5\pm 0.1$ over a dynamic range of $\sim 3$~decades in fluence.  This behavior is similar to the one observed in the SGR bursts \citep{gogus99}. Deviation from the power-law behaviour at the low flux / fluence end of the distribution is due to the limited sensitivity of the ISGRI detector: the weaker bursts visible in the ACS fall below the $3\sigma$ significance threshold which we have adopted for the burst selection in ISGRI. It is interesting to note that if the luminosities of the brightest bursts from \axp\ are $\ge 10^{42}$~erg~s$^{-1}$, the weakest bursts included in our analysis have luminosities  $\sim 10^{39}$~erg~s$^{-1}$, matching the luminosity of the brightest burst from AXPs detected before. Comparing the distribution of the burst fluences in \axp\ with the one of \axpp\ \citep{gavriil04}, we find that in spite of significant differences in the luminosity of individual bursts ($\le 10^{39}$~erg~s$^{-1}$ in \axpp\ vs. $\ge 10^{42}$~erg~s$^{-1}$ in \axp), both distributions have a power-law shape with approximately the same slopes ($0.7\pm 0.1$ in the case of \axpp\ vs. $0.5\pm 0.1$ in \axp).  This indicates that the observed power-law behavior of the  distribution of burst strengths should be valid also for the weaker bursts which are below the ISGRI detection limit. The distribution of the burst peak count rates, shown in the top panel of Fig.~\ref{fig:fluence} has a smaller dynamical range. A powerlaw fit of this distribution, shown in the same panel, is characterized by a slope of $-0.8\pm 0.2$, flatter than that of the distribution of the burst fluxes in \axpp, $1.42\pm 0.13$.

\section{Discussion and conclusions}

We have studied an exceptional period of activity of \axp\ which occurred on 22 January 2009. During this activity period, about $200$ bursts were detected by SPI/ACS and 84 of them were also detected by IBIS/ISGRI detector.  The luminosity of the brightest of these bursts was comparable to the luminosity of the SGR bursts and/or giant flares. This is the first time that bursts of such strength are registered from an AXP. In spite of the fact that the source was outside of the field of view of the {\it INTEGRAL} instruments, we were able to study spectral and temporal characteristics of the bursts, because the signal was so strong that in spite of significant attenuation it was able to penetrate through the walls of the IBIS imager. 

We have found that the activity of the source has exponentially increased on the time scale of about one hour, reaching a peak at around UTC 2009-22-01T06:40. The maximum of the activity was characterized by a very high rate of bursts from the source, with the minimum of the waiting time between subsequent bursts reaching 0.2~s, which is comparable to the typical burst duration. The increase of the burst rate was accompanied by an increase of the strength of the individual bursts. We were not able to measure the flux of the brightest bursts because of saturation effects. However, we estimate the lower  limit on the peak luminosity of the brightest burst to be $L>3\times 10^{42}$~erg~$s^{-1}$. This luminosity is several orders of magnitude larger than that of the previously detected bursts from other AXPs. Our study of the statistical properties of the large amount of the \axp\ bursts reveals that the typical duration of the bursts is $\sim 100$~ms. The distribution of the burst duration extends to  $\sim 10$~s, see Fig.~\ref{fig:duration_histogram}. The typical and maximal burst durations correspond to the time scales of the short spikes and of  the extended afterglows of individual bursts. We find that brighter bursts are characterized by longer durations (Fig.~\ref{fig:flux_duration}) and by harder spectra (Fig.~\ref{fig:hardness_acs}). The extended afterglows of the brightest bursts exhibit pulsations with a period equal to the rotation period of the neutron star. The phases of the maxima of the pulsed afterglow emission of different bursts are close to each other, which indicates that the region (a "trapped fireball" in the magnetosphere, or a "hot spot" on the neutron star surface) is located at the same place in different bursts. At the same time, the bright short spikes of different bursts do not tend to arrive at a preferred phase, which might indicate that the spikes are not produced in the same region as the afterglow emission, as it is expected e.g. in the magnetic reconnection model, in which the bursts occur at high altitudes in the magnetosphere \citep{lyutikov02}. Otherwise, the absence of a correlation with the phase of the pulsed tail emission  could be explained  by the fact that the emission of the bright spikes is relativistically beamed. The beaming effects might also explain the presence of the "orphan" afterglows without spikes and the time delays between the onset of the afterglow and the spike arrival time, observed in a number of \axp\ bursts.

A consistent interpretation of the burst timing and spectral properties observed in the 22 January 2009 activity episode of \axp\ within the two main models of the magnetar flares, the "crustal fracture" and the "magnetic reconnection"  models \citep{thompson95,lyutikov02}, requires further theoretical investigation. It is not clear if the large variety of burst morphologies (which might be somewhat broader than the "type A/B" classification introduced by \citet{woods05}) could be explained within one of the two mechanisms. In spite of the comparable luminosity, it is also not clear if the mechanism of production of the \axp\ bursts is the same as that of the SGR bursts.

Observation of SGR-like bursts from an AXP adds additional argument in favor of the hypothesis that AXPs and SGRs  form  a unique source population, so that \axp\ is an "intermediate" AXP/SGR representative of this population.  Indeed, persistent X-ray emission of \axp\ is characterized by typical AXP spectral characteristics \citep{halpern08}. At the same time \axp\ is distinguished from other AXPs by the presence of pulsed radio emission \citep{camilo07}, which was otherwise only observed in the case of  AXP XTE J1810-197 \citep{camilo06}. At the same time, contrary to most of the SGRs, it appears to have a clear supernova remnant association \citep{gelfand07}. Presence of the "various faces" (radio pulsar / AXP /  SGR) in various states of \axp\ demonstrates that observational properties of AXPs and SGRs are determined by the same physical mechanisms.

\section*{Acknowledgment}

We would like to thank W. Collmar for providing us with access to the ISGRI data for the period of activity of \axp.


\begin{thebibliography}{}
\bibitem[Baldovin et al.(2009)]{baldovin09} Baldovin C., Savchenko V., Beckmann V. et al., 2009, ATEL 1908.
\bibitem[Burgay et al.(2009)]{atel_radio} Burgay N., Israel G. L., Possenti A. et al., 2009, ATEL 1913.
\bibitem[Camilo et al.(2006)]{camilo06} Camilo F.; Ransom S.M., Halpern J.P., Reynolds J., Helfand, D. J.; Zimmerman N., Sarkissian, J., 2006, Nature, 442, 892.
\bibitem[Camilo et al.(2007)]{camilo07} Camilo F., Ransom S.M., Halpern J.P., Reynolds J., 2007, Ap.J., 666, L93.
\bibitem[Camilo et al.(2009)]{atel_radio_no} Camilo F., Halpern J.P., Ransom S.M., 2009, ATEL 1907.
\bibitem[Connaughton \& Briggs(2009)]{gcn_fermi} Connaughton V., Briggs M., 2009, GCN 8835.
\bibitem[Courvoisier et al.(2003)]{courvoisier03} Courvoisier T.J.-L., Walter, R., Beckmann V. et al., 2003, A\& A, 411, L53.
\bibitem[den Hartog et al.(2009)]{denhartog09}  den Hartog P.R., Kuiper L., Hermsen W., 2009, ATEL 1922.
\bibitem[Dib et al.(2008)]{dib08} Dib R., Kaspi V.M., Gavriil F.P., 2008, arXiv:0811.2659.
\bibitem[Duncan \& Thompson(1992)]{duncan92} Duncan R.C., Thompson C., 1992, Ap.J. 392, L9.
\bibitem[Gavriil et al.(2004)]{gavriil04} Gavriil F.P., Kaspi V.M., Woods P.M., 2004, Ap.J., 607, 959.
\bibitem[Gavriil et al.(2002)]{gavriil02} Gavriil F.P., Kaspi V.M., Woods P.M., 2002, Nature, 419, 142.
\bibitem[Gavriil et al.(2007)]{gavriil07} Gavriil F.P., Dib, R., Kaspi V.M., 2007, arXiv:0712.4186. 
\bibitem[Gelfand \& Gaeinsler(2007)]{gelfand07} Gelfand J.D., Gaensler B.M., 2007, Ap.J., 667, 1111.
\bibitem[G\"og\"u\c s et al.(1999)]{gogus99} G\"og\"u\c s E., Woods P.M., Kouveliotou C., van Paradijs J., Briggs M.S., Duncan R.C., Thompson C., 1999, Ap.J., 526, L93. 
\bibitem[G\"og\"u\c s et al.(2001)]{gogus00} G\"og\"u\c s E., Kouveliotou C., Woods P.M., Thompson C., Duncan R.C., Briggs M.S., 2001, Ap.J. 558, 228.
\bibitem[Golenetskii et al.(2009)]{atel_february} Golenetskii S., Aptekar R., Mazets E. et al., 2009, GCN 8913.
\bibitem[G\" otz et al.(2006)]{gotz06} G\" otz D., Mereghetti S., Tiengo A., Esposito P., 2006, A\& A, 449, L31.
\bibitem[Gronwall et al.(2009)]{gcn_swift} Gronwall C., Holland S. T., Markwardt C. B. et al., 2009, GCN 8833.
\bibitem[Halpern et al.(2008)]{halpern08} Halpern J.P., Gotthelf E.V., Reynolds J., Ransom S.M., Camilo F., 2008, Ap.J., 676, 1178.
\bibitem[Hurley et al.(1994)]{hurley94} Hurley K. J., McBreen B., Rabbette M., Steel S., 2004, A\&A, 288, L49.
\bibitem[Hurley et al.(2005)]{nature05} Hurley K., Boggs S. E., Smith D. M. et al., 2005, Nature, 434, 1098. 
\bibitem[Israel et al.(2007)] {israel07} Israel G.L., Campagna S., Dall'Osso S., Muno M.P., Cummings J., Perna R., Stella L., 2007, Ap.J., 664, 448.
\bibitem[Ibrahim et al.(2001)]{ibrahim01} Ibrahim, A.~I., Strohmayer
T.~E., Woods P.~M., Kouveliotou C., Thompson C., Duncan R.~C., Dieters
S., Swank J.~H., van Paradijs J., Finger M., Ap.J, 558, 237
\bibitem[Kaspi et al.(2003)]{kaspi03} Kaspi V.M., Gavriil F.P., Woods P.M., Jensen J.B., Roberts M.S.E., Chakrabarty D., 2003, Ap.J., 588, L93.
\bibitem[Kouveliotou et al.(2009)]{atel_february1} Kouveliotou C., von Kienlin A., Fishman G. et al., 2009, GCN 8915.
\bibitem[Kuiper et al.(2009)]{atel_period} Kuiper L., den Hartog P.R., Hermsen W., 2009, ATEL 1921.
\bibitem[Lamb \& Markert(1981)]{lamb81} Lamb R.C., Markert T.H., 1981, Ap.J., 244, 94.
\bibitem[Lebrun et al.(2003)]{lebrun03} Lebrun F., Leray J. P., Lavocat P. et al., 2003, A\& A, 411, L141.
\bibitem[Lyutikov(2002)]{lyutikov02} Lyutikov M., 2002, Ap.J., 580, L65.
\bibitem[Marcinkowski et al.(2006)]{marcinkowski06} Marcinkowski R., Denis M., Bulik T., Goldoni P., Laurent Ph., Rau A., 2006, A\&A, 452, 113.
\bibitem[Mereghetti et al.(2005)]{mereghetti05} Mereghetti S., G\"otz D., von Kienlin A., Rau A., Lichti G., Weidenspointner G., Jean P., 2005, Ap.J., 624, L105.
\bibitem[Mereghetti(2008)]{mereghetti08} Mereghetti S., 2008, Astron. Astrophys. Rev., 15, 225.
\bibitem[Mereghetti et al.(2009)]{atel2} Mereghetti S., Gotz D., von Kienlin A. et al., 2009, GCN 8841.
\bibitem[Mereghetti et al.(2009a)]{mereghetti09} Mereghetti S., G\"otz D., Weidenspointner G. et al., 2009a, Ap.J.,  696, L74. 
\bibitem[Paczynski(1992)]{paczynski92} Paczynski B., 1992, Acta Astron., 42, 145.
\bibitem[Savchenko et al.(2009)]{atel1} Savchenko V., Beckmann V., Neronov, A. et al., 2009, GCN 8837.
\bibitem[Thompson \& Duncan(1995)]{thompson95} Thompson C., Duncan R.C., 1995, MNRAS, 275, 255.
\bibitem[von Kienlin et al.(2003)]{vonkleinin03} von Kienlin A., Beckmann V., Rau A., et al., 2003, A\& A, 411, L299. 
\bibitem[von Kienlin \& Connaughton(2009)]{gcn_fermi2} von Kienlin A., Connaughton V., 2009, GCN 8838. 
\bibitem[Winkler et al.(2003)]{winkler03} Winkler, C., Courvoisier T. J.-L.; Di Cocco G., et al. 2003, A\&A, 411, L1.
\bibitem[Woods et al.(2005)]{woods05} Woods P.M., Kouveliotou C., Gavriil F.P. et al., 2005, Ap.J., 629, 985.
\end{thebibliography}
  \end{document}